\documentclass[aps,prl,preprintnumbers,showpacs,nofootinbib]{revtex4}
\usepackage[english]{babel}
\usepackage{graphicx}% Include figure files
\usepackage{dcolumn}% Align table columns on decimal point
\usepackage{bm}% bold math
\begin{document}
\newcommand{\newc}{\newcommand}
\newc{\ra}{\rightarrow}
\newc{\lra}{\leftrightarrow}
\newc{\beq}{\begin{equation}}
\newc{\eeq}{\end{equation}}
\newc{\barr}{\begin{eqnarray}}
\newc{\earr}{\end{eqnarray}}
\def \lta {\mathrel{\vcenter
     {\hbox{$<$}\nointerlineskip\hbox{$\sim$}}}}
\def \gta {\mathrel{\vcenter
     {\hbox{$>$}\nointerlineskip\hbox{$\sim$}}}}
\def\vbf{\mbox{\boldmath $\upsilon$}}
\def\barr{\begin{eqnarray}}
\def\earr{\end{eqnarray}}
\def\g{\gamma}
\newcommand{\dphi}{\delta \phi}
\newcommand{\bupsilon}{\mbox{\boldmath \upsilon}}
\newcommand{\at}{\tilde{\alpha}}
\newcommand{\pt}{\tilde{p}}
\newcommand{\Ut}{\tilde{U}}
\newcommand{\rhb}{\bar{\rho}}
\newcommand{\pb}{\bar{p}}
\newcommand{\pbb}{\bar{\rm p}}
\newcommand{\kt}{\tilde{k}}
\newcommand{\kb}{\bar{k}}
\newcommand{\wt}{\tilde{w}}
%%%%%%%%%%%%%%%%%%%%%%%%%%%%%%%%%%%%%%%%%%%%%%%%%%%%%%%%%%%%%%%%%%%%%\vspace{1cm}
\title{DIRECT DETECTION RATES OF DARK MATTER COUPLED TO DARK ENERGY}
\author{N. Tetradis$^{(1)}$, J. D. Vergados$^{(2),(3)}$\thanks{Vergados@cc.uoi.gr}  and Amand Faessler$^{(3)}$ }
\affiliation{$^{(1)}${\it  Department of Physics, University of Athens, Zographou 157 84, Greece,}}
\affiliation{$^{(2)}${\it Theoretical Physics Division, University
of Ioannina, Ioannina, Gr 451 10, Greece}}
\affiliation{$^{(3)}${\it Institute of Theoretical Physics, University of Tuebingen, Tuebingen Germany.}}
%\begin{center}
%{ \Large \bf Detection Rates of
%Dark Matter Coupled to Dark Energy\\}
%\vspace{0.5cm}
%{N. Tetradis$^{(1)}$
% J.D. Vergados$^{(2),(3)}$ and Amand Faessler^{(3)}}
%\\
%\vspace{0.5cm}
%$^1$ {\it Department of Physics, University of Athens, Zographou 157 84, Greece}
%\\
%$^2$ {\it Department of Physics, University of Ioannina, Ioannina 451 10, Greece}
%\\
%\\
%\vspace{1cm}
\begin{abstract}
%\end{center}
We investigate the effect of a coupling between dark matter and dark energy on the rates for the direct
detection of dark matter. The magnitude of the effect depends
on the strength $\kappa$ of this new interaction relative to gravity.
The resulting isothermal velocity distribution for dark matter in galaxy halos is still Maxwell-Boltzmann
(M-B), but the characteristic velocity and the escape velocity are
increased by $\sqrt{1+\kappa^2}$. We adopt a phenomenological approach and consider values of $\kappa$
near unity.
For such values we find that: (i) The (time averaged) event rate increases for light WIMPs, while it is somewhat reduced for WIMP masses larger than 100 GeV.
(ii)
The time dependence of the rate arising from the modulation amplitude is decreased
compared to the standard M-B velocity distribution.
(iii) The average and maximum WIMP energy increase proportionally to $1+\kappa^2$, which, for sufficiently
massive WIMPs, allows the
possibility of designing experiments measuring $\gamma$ rays following nuclear de-excitation.
\end{abstract}
%\abstract{
%\\
%\vspace{1cm}
%}
\pacs{ 95.35.+d, 12.60.Jv}
 %%%%%%%%%%%%%%%%%%%%%%%%%%%%%%%%%%%%%%%%%%%%%%%%%%%%%%%%%%%%%%%%%%%%
\date{\today}
%%%%%%%%%%%%%%%%%%%%%%%%%%%%%%%%%%%%%%%%%%%%%%%%%%%%%%%%%%%%%%%%%%%%%
\maketitle
%PACS numbers:

\section{Introduction}
The combined MAXIMA-1 \cite{MAX}, BOOMERANG \cite{BOOM},
DASI \cite{DASI} and COBE/DMR Cosmic Microwave Background (CMB)
\cite{COBE} observations imply that the Universe is flat \cite{flat01}
%, $\Omega=1.11\pm0.07$
and most of its energy content is exotic \cite{spergel}. These results have been confirmed and improved
by the recent WMAP data \cite{WMAP}. The deduced cosmological expansion is
consistent with  the luminosity distance as a function of redshift of distant supernovae
\cite{supernova1}--\cite{supernova3}.
According to the scenario favored by the observations
there are various contributions to the energy content of our Universe.
The most accessible energy component is baryonic matter, which accounts for
$\sim 5\%$ of the total energy density. A component that has
not been directly observed is cold dark matter (CDM)): a pressureless fluid that is
responsible for the growth of cosmological perturbations through
gravitational instability. Its contribution to the total energy density is
estimated at
$\sim 25\%$. The dark matter is expected to become more abundant
in extensive halos, that
stretch up to 100--200 kpc from the center of galaxies.
%Ongoing experiments are trying to detect the
%dark matter particles in the halo of the Milky Way \cite{cline}.
The component with the biggest contribution to the energy density has
an equation of state similar to that of a cosmological constant and is characterized as dark energy.
The ratio $w=p/\rho$ is negative and close to $-1$.
This component is responsible for $\sim 70\%$ of the total energy density
and induces
the observed acceleration of the Universe \cite{supernova1}--\cite{supernova3}.
The total energy density of our Universe
is believed to take the critical value consistent with spatial flatness.

%Combining the
%the data of these quite precise experiments, crudely speaking, one finds:
%$$\Omega_b=0.05, \Omega _{CDM}= 0.30, \Omega_{\Lambda}= 0.65$$
Since a non-exotic component cannot exceed $40\%$ of the CDM
\cite {Bennett}, there is room for a component consisting of exotic weakly
interacting massive particles (WIMPs).
%  In fact the DAMA experiment ~\cite {BERNA2} has claimed the observation of one signal in direct
%detection of a WIMP, which with better statistics has subsequently
%been interpreted as a modulation signal \cite{BERNA1}.
 Supersymmetry naturally provides candidates for these dark matter constituents
 \cite{goodwit,ellrosz}.
 In the most favored scenario of supersymmetry, the
lightest supersymmetric particle (LSP) can be described as a Majorana fermion, a linear
combination of the neutral components of the gauginos and
higgsinos \cite{goodwit}--\cite{ref2}. In most calculations the
neutralino is assumed to be primarily a gaugino, usually a bino.
% \section{The Essential Theoretical Ingredients  of Direct Detection.}
 Even though there exists firm indirect evidence for a halo of dark matter
 in galaxies from the
 observed rotational curves, it is essential to
detect  such matter directly \cite{goodwit}--\cite{KVprd}.
Until dark matter is actually detected, we will not be able to
exclude the possibility that the rotation curves result from a
modification of the laws of nature as we currently view them.
The direct detection will also
reveal the nature of the constituents of dark matter.

 The
 possibility of direct detection, however, depends on the nature of the dark matter
 constituents. Since the WIMPs are  expected to be very massive  ($m_{WIMP} \gta 30$ GeV)  and
extremely non relativistic with average kinetic energy $ \langle T \rangle \simeq$
50 KeV $\left(m_{WIMP}/ 100\, {\rm GeV} \right)$, they are not likely to excite the nucleus.
As a result, they can be directly detected mainly via the recoiling of a nucleus
($A,Z$) in elastic scattering. The event rate for such a process can
be computed from the following ingredients:
\begin{enumerate}
\item An effective Lagrangian at the elementary particle (quark)
level obtained in the framework of the prevailing particle theory. For supersymmetry
this is achieved as described  in refs. \cite{ref2,JDV96}, for example.
\item A well defined procedure for transforming the amplitude
obtained using the previous effective Lagrangian from the quark to
the nucleon level, i.e. a quark model for the nucleon. This step
in SUSY models is non-trivial, since the obtained results depend crucially on the
content of the nucleon in quarks other than u and d.
%This is
%particularly true for the scalar couplings, which are proportional
%to the quark masses~\cite{Dree}$-$\cite{Chen}, \cite{JDV06} as well as the
%isoscalar axial coupling \cite{JELLIS93,JDV06}.
\item Knowledge of the relevant nuclear matrix elements
\cite{Ress,DIVA00}, obtained with reliable many-body nuclear wave functions. Fortunately, in the case of the scalar
coupling, which is viewed as the most important, the situation is
a bit simpler, as only the nuclear form
factor is needed.
\item Knowledge of the WIMP density in our vicinity and its velocity distribution. Since the
essential input here comes from the rotational curves,  dark matter candidates other than the
LSP (neutralino) are also characterized by similar parameters.
% In most calculations performed up to now
%one employs a Maxwell-Boltzmann velocity distribution with an
%upper velocity cut off put in by hand at the escape velocity $\upsilon_{esc}=2.84 \upsilon_0\simeq 620$ km/s.
\end{enumerate}

In the past various velocity distributions have been considered for the dark matter gas in our galaxy. The most
popular one is the isothermal Maxwell-Boltzmann (M-B) velocity distribution
with $\langle\upsilon ^2\rangle=3v^2_d\simeq 3\upsilon_0^2/2$, where
$v^2_d  =\langle v^2_x \rangle=\langle v^2_y \rangle=\langle v^2_z \rangle$ and $\upsilon_0$ is the velocity of
the sun around the galaxy, i.e. $\upsilon_0\simeq 220$ km/s.
Extensions of the M-B
distribution have also been considered, in particular these that are axially
symmetric with enhanced dispersion in the galactocentric direction
 \cite {Druk,Verg00}. In such
distributions an upper cutoff $\upsilon_{esc}\simeq 2.84 \, \upsilon_0$ is introduced
by hand, in order to eliminate velocities above the escape velocity.

Non-isothermal models have also been considered. Among these one should
mention the ones with late infall of dark matter into the galaxy, i.e caustic rings
 \cite{SIKIVI1}--\cite{Gelmini}, dark matter orbiting the
 Sun \cite{Krauss}, Sagittarius dark matter \cite{GREEN02}.
The velocity distribution has also been obtained in "adiabatic" models employing  the Eddington proposal
\cite{EDDIN}--\cite{VEROW06}. In such an approach, given the density of matter, one can
obtain a mass distribution that depends both on the velocity and
the gravitational potential. Evaluating this distribution in a given point in space, e.g. in our vicinity, one obtains
 the velocity distribution at that point in a self-consistent manner.
Unfortunately this approach is applicable only if the density
of matter is spherically symmetric.

In the present work we will consider another variant of the isothermal M-B distribution, that results when the dark matter interacts
with the dark energy \cite{tetradis,brouzakis}.
The difficulty with explaining the very small value of the
cosmological constant that could induce the present acceleration has
motivated the suggestion that this energy component is time dependent
\cite{wetdil,peeblesold}. In the simplest realization, it is connected to
a scalar field $\phi$ with a very flat potential. The vacuum energy
associated with this field is the dark energy that drives
the acceleration.
If such a field affects
the cosmological evolution today, its effective mass must be of the order
of the Hubble scale, or smaller.

It is conceivable
that there is a coupling between dark matter and
the field responsible for the dark energy \cite{wetcosmon}.
In such a scenario it may be
possible to resolve the coincidence problem, i.e. the reason behind
the comparable present
contributions from the dark matter and the dark energy to the
total energy density.
The presence of an interaction between dark matter
and the scalar field responsible for the dark energy
has consequences that are potentially observable.
The cosmological implications depend on the form of the coupling,
as well as on the potential of the field \cite{amendola}.
If the scale for the field mass is
set by the present value of the Hubble parameter, then
the field is effectively massless at distances of
the order of the galactic scale.
Its coupling to the dark matter particles results in a long range
force that can affect the details of structure formation
\cite{largescale1}--\cite{peeblessim}.

The attraction between
dark matter particles mediated by the scalar field
is expected to modify the
distribution and velocity of dark matter particles in halos, with
implications for dark matter searches.
A careful analysis indicates that the distribution remains Maxwell-Boltzmann, but with a
potentially larger characteristic velocity \cite{tetradis,brouzakis}.
This has two consequences:
\begin{itemize}
\item The total detection rate is reduced for large WIMP masses (above 100 GeV).
This occurs because the velocity distribution is shifted  to higher
values. As a result, such a distribution tends to favor a high energy transfer to the nucleus.
The nuclear form factor tends to
 suppress the high energy transfer components, resulting in an overall suppression.
 \item The modulation effect, i.e. the periodic dependence of the rate on the
Earth's motion, is reduced. This is unfortunate, because the modulation is viewed as a good signature against the background.
\item As the average WIMP velocity increases, the average WIMP energy increases as well. The kinetic energy becomes
\beq
\langle T\rangle  \approx
50(1+\kappa^2) \,{\rm KeV}  \frac{ m_{WIMP}}{ 100\, {\rm GeV}}.
\label{avenergy} \eeq
Thus, for $m_{WIMP} =200$ GeV, one finds $\langle T \rangle \simeq 0.32$ and $1.3$ MeV
for $\kappa^2=1$ and $3$ respectively. Since a value $\kappa^2 \simeq  1$ cannot be
excluded from the available constraints, there is, in this case, a reasonable
possibility for exciting the nucleus. In such a scenario the previous two
disadvantages  are not relevant, as they are connected with nuclear recoil experiments.
This possibility is indeed good news, because measuring the de-excitation
$\gamma$ rays is a much simpler task than the detection of the recoiling nuclei.
\end{itemize}
The above conclusions depend only on the velocity distribution and nuclear structure and are independent
of the specific nature of the WIMP.

It must be emphasized that it is not easy to construct extensions
of the Standard Model that include a dark energy field coupled to dark matter. The main obstruction is
related to the necessity to keep the mass of the field of the order of the present Hubble scale after
radiative corrections. A large coupling to the dark matter field induces significant loop corrections to
the potential of the dark energy field, resulting in a large mass \cite{doran}. On the other hand, it is reasonable to
expect that the resolution of the coincidence problem will require a coupling that is not much smaller than
the gravitational one. For this reason our analysis will be essentially phenomenological. We will assume that
the dark energy field has a mass of the order of the Hubble scale and a coupling to the dark matter of
gravitational strength.  Explicit models that realize these assumptions are given in refs. \cite{mainini,massimo}.

\section{Interaction between dark matter and dark energy}

We consider an interaction between the scalar field and the
dark matter particles that can be modeled through
a field-dependent particle mass. The action takes
the form
\begin{equation}
{\cal S}=\int d^4x \sqrt{-g}
\left(M^2 R -\frac{1}{2}g^{\mu\nu}
{\partial_\mu \phi}\,
{\partial_\nu \phi}
-U(\phi) \right)
-\sum_i \int m(\phi(x_i))d\tau_i,
\label{one} \end{equation}
where $d\tau_i=\sqrt{
-g_{\mu\nu}(x_i)dx^\mu_idx^\nu_i}$ and
the second integral is taken over particle trajectories.
Variation of the action with respect to $\phi$ results in the
equation of motion
\beq
\frac{1}{\sqrt{-g}}{\partial_\mu}
\left(\sqrt{-g}\,\,g^{\mu\nu}{\partial_\nu \phi}
\right)=
\frac{dU}{d\phi}-\frac{d\ln m(\phi(x))}{d\phi}\,\, T^\mu_{~\mu},
\label{two} \eeq
where the energy-momentum tensor associated with the gas of particles
is
\beq
T^{\mu\nu}=\frac{1}{\sqrt{-g}}
\sum_i \int d\tau_i \,\, m(\phi(x_i))\,\,
\frac{dx_i^\mu}{d \tau_i}\frac{dx_i^\nu}{d \tau_i}
\delta^{(4)}(x-x_i).
\label{three} \eeq

We are interested in static spherically symmetric configurations,
with the scalar field varying slowly with the radial distance $r$.
Our treatment is relevant up to
a distance $r_1 \sim 100$ kpc beyond which the dark matter becomes very
dilute. For $r \gta r_1$
we expect
that $\phi$ quickly becomes constant with a value close
to $\phi(r_1)\equiv \phi_1$.
This is the value that
drives the present cosmological expansion. Here we assume that the
cosmological evolution of
$\phi_1$ is negligible for the time scales of interest, so that
the asymptotic configuration is static to a good approximation.

We approximate:
$m(\phi)\simeq m(\phi_0)+[dm(\phi_0)/d\phi]\, \dphi
\equiv m_0+m'_0 \, \dphi$, with $\phi_0$ the value of the field at
the center of the galaxy ($r=0$). We work within the
leading order in $\dphi$ and assume that $m'/m\simeq m'_0/m_0$
for all $r$. Also
$dU/d\phi$ can be approximated by a constant between $r=0$ and $r=\infty$.
For the scalar field to provide a resolution of the coincidence problem,
the two terms in the r.h.s. of Eq. (\ref{two}) must be of
similar magnitude in the cosmological solution.
This means that
$dU/d\phi$ must be comparable to $(m'_0/m_0)\rho_\infty$.
We expect $\rho_\infty$ to be a fraction of the critical density, i.e.
$\rho_\infty \sim 3$ keV/cm$^3$.
On the other hand, the energy density
in the central region of the static
solution ($r \lta 100$ kpc) is that of the galaxy halo
($\sim 0.4$ GeV/cm$^3$ for our neighborhood of the Milky Way).
This makes $dU/d\phi$ negligible in the r.h.s. of
Eq. (\ref{two}) for a static configuration. The potential is
expected to become important only for $r \to \infty$, where the
static solution must be replaced by the cosmological one.
Similar arguments indicate that we
can neglect $U$ relative to $\rho$. Also the scalar field must be effectively
massless at the galactic scale.
For these reasons we expect that the form of the
potential plays a negligible role at the galactic level. Our analysis
can be carried out with $U=0$ and is model independent.

We treat
the dark matter as a weakly interacting, dilute gas.
We are motivated by the phenomenological
success of the isothermal sphere \cite{peeblesb} in
describing the flat part of the rotation curves.
We do not address the question of the density profile in the inner part of
the galaxies ($r \lta 5$ kpc).
We approximate the energy-momentum tensor of the dark matter
as $T^\mu_{~\nu}={\rm diag} (-\rho,p,p,p)$ with
$p(r)=\rho(r)\, \langle v_d^2 \rangle
=m(\phi(r))\, n(r)\, \langle v_d^2 \rangle$. The
dispersion
%of each component
of the dark matter velocity is
assumed to be constant and small: $\langle v_d^2 \rangle \ll 1$.
The gravitational field is considered in the
Newtonian approximation: $g_{00}\simeq 1+2\Phi$, with
$\Phi = {\cal O}\left( m'_0\dphi/m_0\right)$.
In the weak field limit and for
$p \ll \rho$, the conservation of the energy-momentum tensor gives
\beq
p'=-\rho\, \Phi'-\rho \, \frac{m'_0}{m_0} (\dphi)',
\label{extra} \eeq
with the prime on $p$, $\Phi$, $\dphi$ denoting a derivative with
respect to $r$. Integration of this equation gives
$n\simeq n_0
\exp \left(-\Phi/\langle v_d^2 \rangle
-(m'_0/m_0)\dphi/\langle v_d^2 \rangle \right).$
%We can define an effective temperature $T_0$
%through the relation:
%$\langle v_d^2 \rangle=T_0/m_0$.
%In order to make the picture more complete we also
%allow for a pressureless baryonic component in the core. We assume that its
%energy density has the phenomenological profile
%$\rho_b(r)=\rho_B\, f \left( r/r_c \right)$, with $f(x)$ a
%decreasing function of $x$.
%We do not discuss the physics that leads to such a profile.
%We include the baryonic contribution only in order to
%estimate its effect on the dark matter distribution in the
%outer regions of the galaxy.

With the above assumptions we obtain the equations of motion
\beq
\Phi''+ \frac{2}{r}\Phi'=\frac{1}{4M^2} \, \rho_0
\exp \left(-\alpha \Phi-\at \dphi \right),
\label{five} \eeq
and
\beq
(\dphi)''+\frac{2}{r}(\dphi)'=\frac{m'_0}{m_0}\, \rho_0
\exp \left(-\alpha \Phi-\at \dphi \right),
\label{six} \eeq
where $M=(16\pi G_N)^{-1/2}$ is the reduced Planck mass,
$\rho_0=m_0n_0$ the energy density of dark
matter at $r=0$, $\alpha
%=m_0/T_0
=1/\langle v_d^2 \rangle$,
and $\at=m'_0/(m_0 \langle v_d^2 \rangle)$.
We emphasize that, even though $|\Phi| \ll 1$, the
combination $\Phi/ \langle v_d^2 \rangle$,
that appears in the exponent in the
expression for the number density $n$, can be large.
Similarly, the expansion of the mass around the value
$m_0=m(\phi_0)$ assumes the smallness of the dimensionless parameter
$|m'_0 \dphi/m_0|$. However, the combination
$\at \dphi=(m'_0 \dphi/m_0)/\langle v_d^2 \rangle$,
that appears in the exponent, can be
large.

A linear combination of Eqs. (\ref{five}), (\ref{six}) gives
\beq
\frac{d^2u}{dz^2}+\frac{2}{z}\frac{du}{dz}+\exp u,
\label{seven} \eeq
where
$u=-\alpha \Phi-\at \dphi$, $z=\beta r$
and $\beta^2=(1+\kappa^2)\alpha {\rho_0}/{4M^2}$.
The parameter
\beq
\kappa^2=4M^2\left( {m'_0}/{m_0}\right)^2
\label{kappa2} \eeq
determines the strength of the new interaction relative to gravity.
The solutions that are regular for small  $z$ approach the form
\beq
u=\ln\left(\frac{2}{z^2}\right)+\frac{1}{\sqrt{z}}
\left[
d_1\cos\left(\frac{\sqrt{7}}{2}\ln z \right)
+d_2\sin\left(\frac{\sqrt{7}}{2}\ln z \right)
\right]+...
\label{ucor} \eeq
for large $z$.
Another linear combination of Eqs. (\ref{five}), (\ref{six}) gives
\beq
\frac{d^2w}{dz^2}+\frac{2}{z}\frac{dw}{dz}=0,
\label{eight} \eeq
with
$w=-\kappa^2 \alpha \Phi+\at\dphi.$
The solution of this equation is
$w=c_0+{c_1}/{z}.$

The velocity $v$ of a massive baryonic object in orbit around the galaxy,
at a distance $r$ from
its center, can be expressed as
\beq
\left(\frac{v}{v_c}\right)^2=\frac{r\Phi'}{v^2_c}=-\frac{z}{2}
\left(\frac{du}{dz}+\frac{dw}{dz} \right),
\label{rotvel} \eeq
where
\beq
v_c^2=\frac{2}{1+\kappa^2}
\langle v^2_{d} \rangle.
\label{vel} \eeq
The asymptotic form of $u(z)$, $w(z)$
indicates that $v \simeq v_c$ for large $z$. The
dominant correction to the leading behavior
arises from the term $\sim 1/\sqrt{z}$ in Eq.
(\ref{ucor}). The function $v(z)$ gives a higher order correction.
%This means that the presence of the
%field $\phi$ is not expected to cause significant modifications
%to the shape of the rotation curves
%relative to the $\phi=0$ case.
This simple analysis indicates that
the approximately flat rotation curves outside the galaxy cores
are a persistent feature
even if the dark matter is coupled to a scalar field through
its mass.
%However, numerical simulations are probably necessary
%in order to reproduce
%the detailed form of the curves.
If the new interaction is universal for ordinary and dark matter, the
experimental constraints impose $\kappa^2 \ll 1$. In this case, it is
reasonable to expect a negligible effect in the distribution of
matter in galaxy halos.
However, if $\phi$ interacts only with dark matter, as we assume here,
this bound can
be relaxed significantly.

A massive particle in orbit around the galaxy, at a large distance $r$ from
its center, has a velocity given by Eq. (\ref{vel}).
We can use this expression in order to fix
$\langle v^2_d \rangle$ for
a given value of $\kappa$.
The effect of the new scalar interaction is encoded in the factor
$\kappa^2$. When this is small, the velocity of
an object orbiting the galaxy is of the order of the square root of
the dispersion of the
dark matter velocity. If $\kappa^2$ is large, the
rotation velocity can become much smaller than the typical
dark matter velocity.

The allowed range of $\kappa$ is limited by the observable implications
of the model that describes the dark sector.
It is reasonable to expect that
the resolution of the coincidence problem through an interaction
between dark matter and dark energy will
have to rely on a coupling not significantly
weaker that gravity. It seems unlikely that a coupling $\kappa^2\ll 1$
can lead to a cosmological evolution drastically different from
that in the decoupled case.

The dependence of the mass of
dark matter particles on an evolving scalar field
during the cosmological evolution since the
decoupling is reflected in the microwave background. The magnitude of the
effect is strongly model dependent. In the models of ref.
\cite{amendola,mainini} the observations result in
the constraint $\kappa^2 \lta 0.01$.
In the model of ref. \cite{peeblessim} the
scalar interaction among dark matter particles is screened by an additional
relativistic dark matter species. As a result,
the model is viable even for couplings
$\kappa^2 \simeq 1$.
A similar mechanism is employed in ref. \cite{massimo}.
In this model
the interaction between dark matter and dark energy
becomes important only during the recent evolution
of the Universe. In general, an interaction that is effective for
redshifts $z \lta 1-2$ is not strongly constrained by the
observations.

Independently of the value of $\kappa^2$, the interaction of dark matter
with the scalar field associated with dark energy does not destroy the
approximately
flat profile of the rotation curves. Other considerations, however, could
constrain the coupling
$\kappa^2$.
The dispersion of the dark matter velocity is
$\langle v^2_{d} \rangle=(1+\kappa^2) v_c^2/2$.
For a value of $v_c$ deduced from observations,
$\langle v^2_{d} \rangle$ increases with $\kappa$. For sufficiently
large $\kappa$, it seems possible that $v_c$
may exceed the escape velocity from the galaxy.
It turns out, however, that this is not the case. Outside the core of the
galaxy and for
$r\lta r_1$, the binding potential for
a dark matter particle is $\Phi+(\at/\alpha)\dphi$. For large $r$,
Eq. (\ref{eight}) implies that $v=-\kappa^2\alpha\Phi+\at \dphi=$ constant. The
binding potential becomes $(1+\kappa^2)\Phi=(1+\kappa^2)v^2_c\ln(r/r_1)$,
where we have omitted an overall constant. For a particle
at a distance $r_*$ from the center of the galaxy,
the escape velocity becomes
\beq
v^2_{esc}=2(1+\kappa^2)[\ln(r_1/r_*)+1].
\label{escv} \eeq
The value of $v_{esc}$
is larger
than the standard one \cite{spergel}
by a factor $(1+\kappa^2)$,
so that $\langle v^2_{d} \rangle$ remains substantially
smaller than $v^2_{esc}$ for $r_* \ll r_1$.
A particle that does not interact with the
scalar field is bound only by the the potential $\Phi$. However, the scale of
its velocity is set by $v_c$, so that again it cannot escape.

\section{The velocity distribution of dark matter}
In the previous section we saw that in isothermal models
 the dark matter velocity distribution with respect to the galactic center is M-B:
\begin{equation}
f(v)=\frac{1}{\pi \sqrt{\pi}}\frac{1}{v_{m} ^3} \exp\left(-\frac{v^2}{v_m ^2}\right),
\end{equation}
where $v^2_m =2 v^2_d=(1+\kappa^2)v_c^2$, with $ v_c$ the observed rotation
velocity of a baryonic object in orbit around the galaxy. This means that the dispersion of the dark matter velocity
is proportional to $1+\kappa^2$, where $\kappa$ is the coupling between dark matter and dark energy, given
by Eq. (\ref{kappa2}).
We have also assumed that the
ordinary baryonic matter does not couple to the dark energy field and is not affected by its presence.
%$ \upsilon_m=\sqrt{\frac{2}{3}\prec \upsilon^2\succ }$ is a characteristic velocity, which is depends on the temperature.
We impose an upper bound
$v_b$ on the dark matter velocity, equal to the escape velocity given by Eq. (\ref{escv}).
We express it as
\beq
v_b=n\, v_{esc,0},
\label{escvv} \eeq
where $v_{esc,0}$ is the escape velocity for $\kappa=0$ and $n^2=1+\kappa^2$.

In the local frame this velocity distribution takes the form
%$f(\mbox{\boldmath $\upsilon$},\mbox{\boldmath $\upsilon$}_E)$
%$f(\mbox{\boldmath $\upsilon$})$
%$$f(\mbox{\boldmath $\upsilon$})$$
\begin{equation}
f({\vbf})=\frac{1}{\pi \sqrt{\pi}} \frac{1}{v_m^3} \exp\left( -\frac{\left({\vbf}+{\vbf_E}\right)^2}{v^2_m} \right),
\label{distrlocal} \end{equation}
where
\beq
\mbox{\boldmath $\upsilon$}_E  = \mbox{\boldmath $\upsilon$}_0  +  v_1
(\sin{\alpha} \,\, {\bf \hat x}
-\cos {\alpha} \, \cos{\gamma} \, \,{\bf \hat y}
+ \cos {\alpha} \, \sin{\gamma} \, \,{\bf \hat z} \,).
\label{param} \eeq
We have chosen a coordinate system in which the polar z-axis is  along the direction of motion of the Sun,
the x-axis is radially out of the galaxy and ${\bf \hat{y}}={\bf \hat{z}}\times {\bf \hat{x}}$.
% ${\bf v}_0$ is the velocity of the sun around the center of the galaxy and
The velocity of the Earth is $\mbox{\boldmath $\upsilon$} _E$.
The
velocity of the Sun around the center of the galaxy is $\upsilon_0 \,{\bf \hat{z}}$
(with $\upsilon_0\simeq$ 220 km/s).
The magnitude of the velocity of the Earth relative to the Sun is $\upsilon_1\simeq 30$ km/s.
The quantity $\gamma\simeq \pi/6$ describes the orientation of the ecliptic
with respect to the galactic plane. (The angle between the normals to the two planes is $\pi/2-\gamma\simeq \pi/3$.)
The parameter $\alpha$ denotes the phase of the Earth. ($\alpha=0$ on June 2nd.)

In the standard scenario, in which there is no interaction between dark matter and dark energy,  we have $\upsilon_m=\upsilon_0$ and
$\upsilon_b=y_{esc} \upsilon_0$ with $y_{esc}\simeq 2.84$.
In the scenario we are considering both parameters are scaled up by the same factor,
i.e. $\upsilon_m= n\, \upsilon_0$ and $\upsilon_b=n\,y_{esc}\,  \upsilon_0$, with
$n=\sqrt{1+k^2}$. The standard M-B distribution is
a special case of our model with $n=1$. In the present work we treat $n$ as a free parameter, which we do not expect to be
much larger than unity.  We will consider values as large as
$n=2$ (that corresponds to $\kappa=\sqrt{3}$) and study the implications for direct dark matter detection.
The distribution function can be written as:
\beq
f(y,\xi,\phi,\delta,n)=\frac{1}{\pi  \sqrt{\pi}} \frac{1}{n^3}e^\frac{-y^2-2 \left((2\delta) ^2+y \sqrt{1-\xi ^2} \cos (\phi ) \sin (\alpha )2\delta +\cos
   (\alpha ) \left((y \xi +2) \sin (\gamma )-y \sqrt{1-\xi ^2} \cos (\gamma ) \sin (\phi
   )\right) 2\delta +y \xi +1\right)}{n^2},
   \eeq
where $\phi$ is the azimuthal angle, $\xi$ the cosine of the angle between
$\mbox{\boldmath $\upsilon$} $ and $\mbox{\boldmath $\upsilon$} _0$,
$y={v}/{v_0}$, $\delta=\sin{\gamma} {\upsilon_1}/{\upsilon_0}\simeq({1}/{2})({30}/{220})= 0.068$.
The integral over $\phi$ can be done analytically to yield:
  \beq
\tilde{f}(y,\xi,\delta,n)=\frac{2}{  \sqrt{\pi}} \frac{1}{n^3}e^\frac{-y^2-2 \left((2\delta) ^2 +\cos
   {\alpha } \left((y \xi +2) \sin {\gamma }\right) 2\delta +y \xi +1\right)}{n^2}
I_0\left(2 \delta y \sqrt{(1-\xi^2) (1-\cos^2 {\alpha} \sin^2{\gamma})}\right),
\label{fdis}
   \eeq
%\beq
%\tilde{f}(y,\xi,\delta,n)=\frac{2}{  \sqrt{\pi}} \frac{1}{n^3}e^\frac{-y^2-2 %%\left((2\delta) ^2 +
%\cos(\alpha ) \left((y \xi +2) \sin (\gamma )\right) 2\delta +y \xi +1\right)}
%{n^2}} I_0(2 \delta y \sqrt{(1-\xi^2) (1-\cos^{\alpha} \sin^2{\gamma})})
%\label{fdis}
%\eeq
where $I_0(x)$ is the well known modified Bessel function.
 The various
variables are constrained by:
\beq
\sqrt{y^2+2 \left((2\delta) ^2 +\cos
   \alpha  \left(y \xi +2 \right) \delta +y \xi +1\right)}
\leq n y_{esc}.
\label{condition}
\eeq
From the kinematics of the WIMP-nucleus collision we find that the momentum transfer to the nucleus is given by
\beq
q=2 \mu_r \upsilon \cos{\theta}, \eeq
where $\theta$ is the angle between the WIMP velocity and the momentum of the outgoing nucleus, and
$\mu_r$ the reduced mass of the system. Instead of the angle $\theta$ one can introduce
the energy $Q$ transferred
to the nucleus, $Q={q^2}/({2 A m_p})$ ($A m_p$ is the nuclear mass). Thus
$$2 \sin{\theta} \cos{\theta}d\theta=-\frac{A m_p}{2 (\mu_r \upsilon)^2} dQ.$$
Furthermore, for a given energy transfer the velocity $\upsilon$ is constrained to be
\beq
\upsilon\geq \upsilon_{min}~,~\upsilon_{min}= \sqrt{\frac{ Q A m_p}{2}}\frac{1}{\mu_r}.
\eeq
We will find it convenient to introduce, instead of the energy transfer, the dimensionless quantity $u$
\beq
u=\frac{1}{2}(qb)^2\equiv\frac{Q}{Q_0}~~,~~Q_{0}=\frac{1}{Am_p b^2}=4.1\times 10^{4}~A^{-4/3}~{\rm keV},
\label{u.1}
\eeq
where $b$ is the nuclear (harmonic oscillator) size parameter.

It is clear that for a given energy transfer the velocity is restricted from below. We have already mentioned that the velocity is bounded from above by the escape velocity. We thus get
\beq
a \sqrt{u}\leq y\leq n y_{esc}~,~a= \left[\sqrt{2}\mu_r b \upsilon_0 \right ]^{-1},
\eeq
\beq
2 \sin{\theta} \cos{\theta} d \theta=-\frac{a^2}{y^2} dy.
\label{utrans}
\eeq

\section{The direct detection event rate}
The event rate for the coherent WIMP-nucleus elastic scattering is given by \cite{Verg01,JDV03,JDV04,JDV06}:
\beq
R= \frac{\rho (0)}{m_{\chi^0}} \frac{m}{m_p}~
              \sqrt{\langle v^2 \rangle } f_{coh}(A,\mu_r(A)) \sigma_{p,\chi^0}^{S}
\label{fullrate}
\eeq
with
\beq
f_{coh}(A, \mu_r(A))=\frac{100\mbox{GeV}}{m_{\chi^0}}\left[ \frac{\mu_r(A)}{\mu_r(p)} \right]^2 A~t_{coh}\left(1+h_{coh}cos\alpha \right)
\eeq
In the above expression $\sigma_{p,\chi^0}^{S}$ is the WIMP-nucleon scalar cross section, $\rho(0)$ the WIMP density in our vicinity, $m_{\chi^0}$ the WIMP mass , $m$ the target mass, $A$ the number of nucleons
in the nucleus and $\langle v^2 \rangle=3v_0^2/2$ the average value of the square of the WIMP velocity for
$n=1$.
 The number of events in time $t$ is:
\beq
 R\simeq 1.60~10^{-3}
\frac{t}{1 \mbox{y}} \frac{\rho(0)}{ {\mbox0.3GeVcm^{-3}}}
\frac{m}{\mbox{1Kg}}\frac{ \sqrt{\langle
v^2 \rangle }}{280 {\mbox kms^{-1}}}\frac{\sigma_{p,\chi^0}^{S}}{10^{-6} \mbox{ pb}} f_{coh}(A, \mu_r(A))
\label{eventrate}
\eeq
%\beq
%R\simeq 160~10^{-4}~({\rm pb})^{-1} y^{-1}\frac{\rho(0)}{0.3{\rm GeV} {\rm cm}^{-3}}\frac{m}{1{\rm Kg}}\frac{ %\sqrt{\langle v^2 \rangle }}{280\,{\rm km}\, {\rm s}^{-1}}
%f_{coh}(A,\mu_r(A)) \sigma_{p,\chi^0}^{S},
%\label{eventrate}\eeq
%with
%\beq
%f_{coh}(A, \mu_r(A))=\frac{100\mbox{GeV}}{m_{\chi^0}}\left[ \frac{\mu_r(A)}{\mu_r(p)} \right]^2 A~t_{coh}\left(1+h_{coh}\cos\alpha \right).
%\eeq
The quantity of interest to us is $r= t_{coh}\left(1+h_{coh}\cos\alpha \right)$, which contains all the information
regarding the WIMP velocity distribution and the structure of the nucleus. It also depends on the reduced mass of the system.

The event rate is proportional to the WIMP flux, i.e. proportional to the WIMP velocity.
In Eq. (\ref{eventrate}) we have chosen to normalize the
event  rate using the velocity dispersion for $n=1$, i.e.
 $\sqrt{\langle \upsilon^2\rangle }=\sqrt{{3}/{2}} \upsilon_0$. As a result a compensating factor of
${\upsilon}/{\sqrt{\langle \upsilon^2\rangle }}$ is included in $r$.
It is not difficult to show  \cite{Verg01,JDV03,JDV04,JDV06} that
\beq
\frac{dr}{du}= F^2(u) \int_{a \sqrt{u}}^{yesc} \sqrt{\frac{2}{3}} y \frac{a^2}{y^2} y^2 dy \int_{-1}^{\xi_0(y,\alpha )}
\tilde{f}(y,\xi,\delta,n)d \xi,
\eeq
where $F(u)$ is the nuclear form factor. In the integrand we have displayed explicitly all the factors of y
in order
to keep track of their origin. The first one comes from the flux,
the second from the transformation (\ref{utrans}) and the last is the usual
phase-space factor. The quantity $\xi_0(y,\alpha )$ enters because in some region of the velocity space
the upper value of $\xi$ is restricted so that the condition (\ref{condition}) is satisfied. The above expression can be cast in the form:
\begin{equation}
\frac{dr}{du}=\sqrt{\frac{2}{3}} a^2 F^2(u)  \Psi(a \sqrt{u},\alpha ).
\label{eq:rrateall}
\end{equation}
We have seen that the parameter $a$ depends on the nucleus, $\upsilon_0$ and the WIMP mass.

In spite of the complications arising from the condition (\ref{condition}), by taking the leading order of the modified Bessel function
in Eq. (\ref{fdis}), we were able to get an analytic
expression for $\Psi(x,\alpha )$ as follows:
\beq
\Psi(x,\alpha)=(1-\Theta(x-y_{esc}+1+\delta \cos{\alpha})) \Psi_{<}(x,\alpha)+
\Theta(x-y_{esc}+1+\delta \cos{\alpha}) \Psi_{>}(x,\alpha)
\eeq
\barr
\Psi_{<}=(x,\alpha)&=&
\frac{-{\rm erf}\left(\frac{y_{esc}}{n}\right)+
{\rm erf}\left(\frac{-x+\delta  \cos {\alpha
   }+1}{n}\right)+\text{erf}\left(\frac{x+\delta  \cos
   {\alpha }+1}{n}\right)-
{\rm erf}\left(\frac{2 (\delta  \cos {\alpha }+1)-y_{esc}}{n}\right)}{2 (\delta
   \cos {\alpha }+1)}\\
\nonumber
&+&{\rm erf}\left(\frac{2   (\delta  \cos {\alpha}+1)-y_{esc}}{n}\right)
-\frac{{\rm erf}\left(\frac{-y_{esc}+\delta \cos {\alpha }+1}{n}\right)}{2 (\delta  \cos {\alpha
   }+1)}-\frac{e^{-\frac{y_{esc}^2}{n^2}}}{n \sqrt{\pi }}
\earr
\beq
\Psi_{>}=(x,\alpha)=
\frac{\frac{2 e^{-\frac{y_{esc}^2}{n^2}} (x-y_{esc})}{n \sqrt{\pi
   }}-{\rm erf}\left(\frac{-y_{esc}+\delta  \cos {\alpha
   }+1}{n}\right)+{\rm erf}\left(\frac{-x+\delta  \cos
   {\alpha }+1}{n}\right)}{2 (\delta  \cos {\alpha
   }+1)},
\eeq
where $x$ is a short hand notation for $a\sqrt{u}$ and  $\Theta(x)$ is the Heavyside
step function.

%Integrating $y f(y) $ from $x$ to a maximum 2.84, we obtain the
%effect of the velocity distribution on the average rate $\Psi(x)$ and the modulation $H(x) \cos{\alpha}$. In
%practice $x=a \sqrt{u}$, a being a parameter depending on the LSP mass and the nucleus and u  the energy transferred to the nucleus
%in some convenient dimensionless units. The functions $\Psi(a \sqrt(u)$ and the modulation $H(a \sqrt{u})$   give
%the effect of the velocity distribution on the rate. The differential rate is proportional to the quantity:
By performing a Fourier analysis of the function $\Psi(x,\alpha)$, which is a periodic function of $\alpha$,
 and keeping the dominant terms we find:
\begin{equation}
\frac{dr}{du}= \sqrt{\frac{2}{3}} a^2 F^2(u)\left [ \Psi_0(a \sqrt{u})+H(a \sqrt{u}) \cos{\alpha}+H_2(a \sqrt{u}) \cos{2 \alpha} \right ].
\label{eq:rrate}
\end{equation}
Sometimes we will consider separately each term in the above expression by writing:
\beq
\frac{dr}{du}= \frac{dt}{du}+\frac{dh}{du}\cos{\alpha}+\frac{dh2}{du}\cos{2 \alpha}.
\eeq
%The factor $n \sqrt{\frac{2}{3}}$ has been introduced to comply with the usual calculations, which employ the standard M-B
%quantity $\sqrt{\prec \upsilon^2\succ}$ in the flux. The quantity $a$ depends on the reduced mass $\mu_r$ of the
%LSP-nuclear system, the size of the nucleus $b$ and the parameter $n$ of the M-B distribution and is given by:
%\beq a= \left[\sqrt{2}\mu_r b n \upsilon_0 \right ]^{-1}, \label{a.1} \eeq
%The quantity $u$ is related to the energy transfer $Q$ via the relation:
%\beq
%u=\frac{Q}{Q_0}~~,~~Q_{0}=4.1\times 10^{4}~A^{-4/3}~KeV
%\label{u.1}
%\eeq

\begin{figure}[t]
\begin{center}
\includegraphics[clip,width=1.0\linewidth]{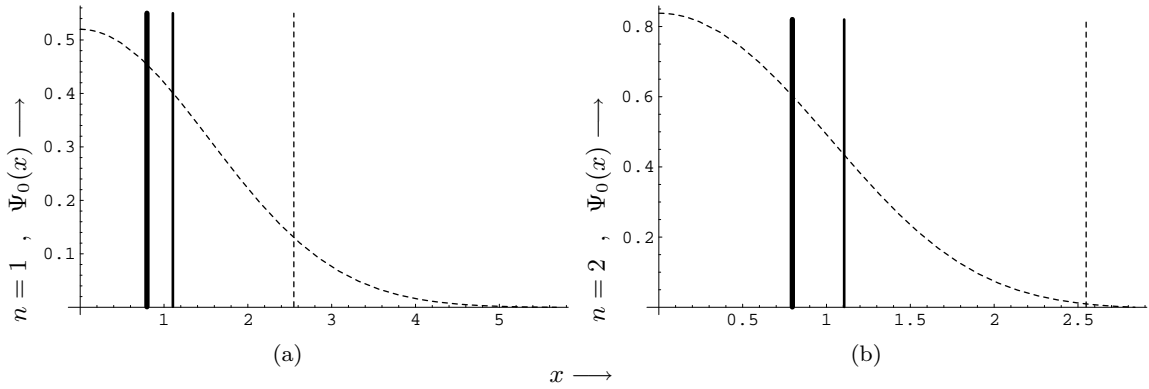}
\caption{The function $\Psi_0(x)$ as defined in the text. From left to right $n=1$ and 2. The area
 under the curve is roughly independent of $n$.
Due to the nuclear form factor, not all the range of $u$ is exploitable in direct WIMP detection. For $^{127}I$there is
effectively a cut off value indicated by a dotted line, a fine line and a thick line for a WIMP mass of $30$, $100$ and $200$ GeV respectively. The exploitable area under the curve decreases as $n$ increases.}
 \label{fig:nomod}
 \end{center}
  \end{figure}

\begin{figure}[!ht]
 \begin{center}
 \includegraphics[clip,width=1.0\linewidth]{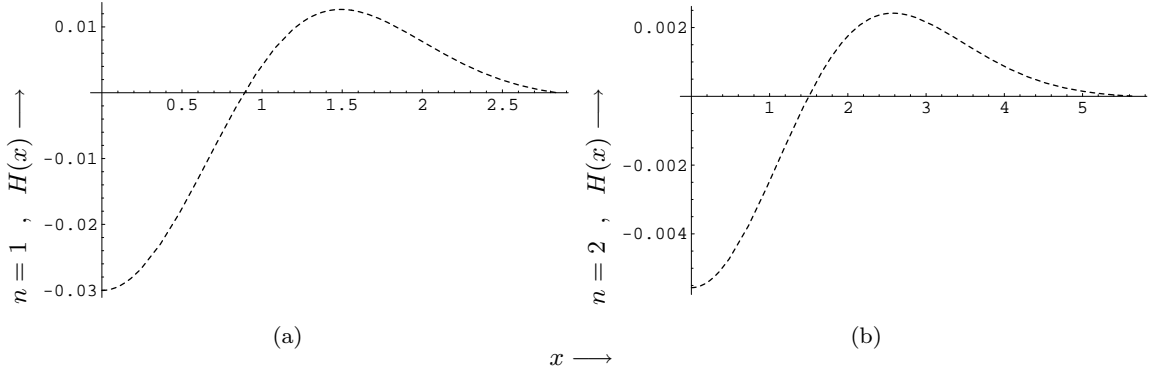}
 \caption{The function $H(x)$ giving the effect of the velocity distribution on the modulated differential rate. From left to right $n=1$ and 2. Note the change in sign and the fact that the amplitude decreases with incensing n.}
 \end{center}
 \label{fig:mod}
  \end{figure}

\begin{figure}[!ht]
 \begin{center}
 \includegraphics[clip,width=1.0\linewidth]{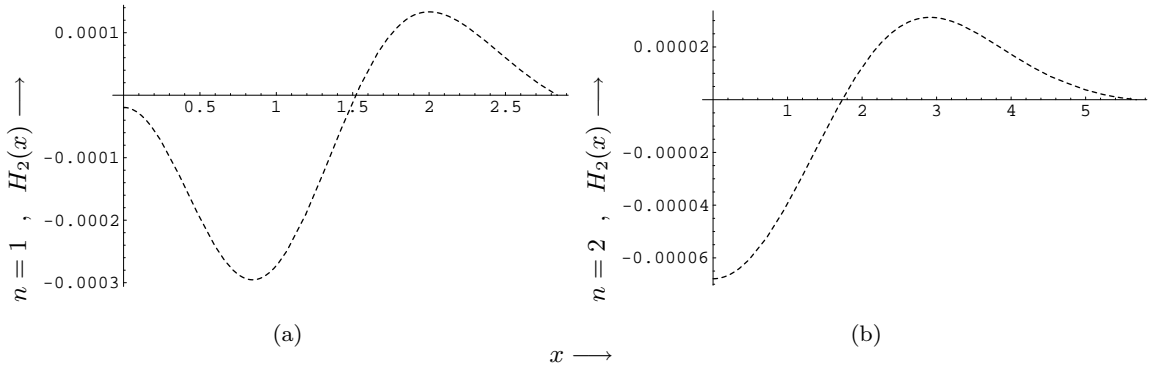}
 \caption{The function $H_2(x)$ giving a higher order effect on the modulated differential rate. From left to right $n=1$ and 2. Note again the change in sign and the fact that the amplitude decreases with increasing n.}
 \label{fig:mod2}
 \end{center}
  \end{figure}

Before proceeding further by considering a special target, it is instructive to concentrate on the dependence of $\Psi_0(x)$ and the modulation $H(x)$
on the parameter $n$ of the M-B distribution. For this purpose we exhibit the function $\Psi_0(x)$
in Fig. \ref{fig:nomod}, the function $H(x)$
in Fig. \ref{fig:mod} and $H_2(x)$
in Fig. \ref{fig:mod2}.
From Fig. \ref{fig:nomod} it is apparent that the high energy transfers are cut off because of the nuclear form
factor at values lower than the limit imposed by the upper bound on the WIMP velocity, which increases with $n$.
In the case of  $\Psi_0(x)$, one clearly sees that the peak value decreases with $n$.
Even though the area under the curve remains roughly independent of $n$, the portion available to direct detection decreases because of the nuclear
form factor.
 In the case of $H(x)$ one notices a change in sign. The low $Q$ section tends to cancel the high $Q$ part, when one
integrates over $Q$
  to get the total event rate. Furthermore we see that the modulation amplitude is decreasing with $n$.
  By comparing $H(x)$ and $H_2(x)$, one can see that the effect of the higher Fourier components in $\alpha$ is negligible.
  \section{Applications}
  As we have already mentioned, the absolute rate depends critically on the specific nature of the WIMP, e.g. on the SUSY parameters
in the case of the neutralino. It also depends on the structure of the nucleon.  In the present work we will not be concerned
with those very important aspects (see e.g. Refs \cite{Verg01,JDV03,JDV04,JDV06} on how one deals with such issues). The event rate is proportional to the WIMP density in our vicinity, which is not modified by including the coupling between dark matter and dark energy as in our model. In any case, we will focus here on the aspects affected by the WIMP velocity distribution.

\begin{figure}[!ht]
 \begin{center}
 \includegraphics[clip,width=1.0\linewidth]{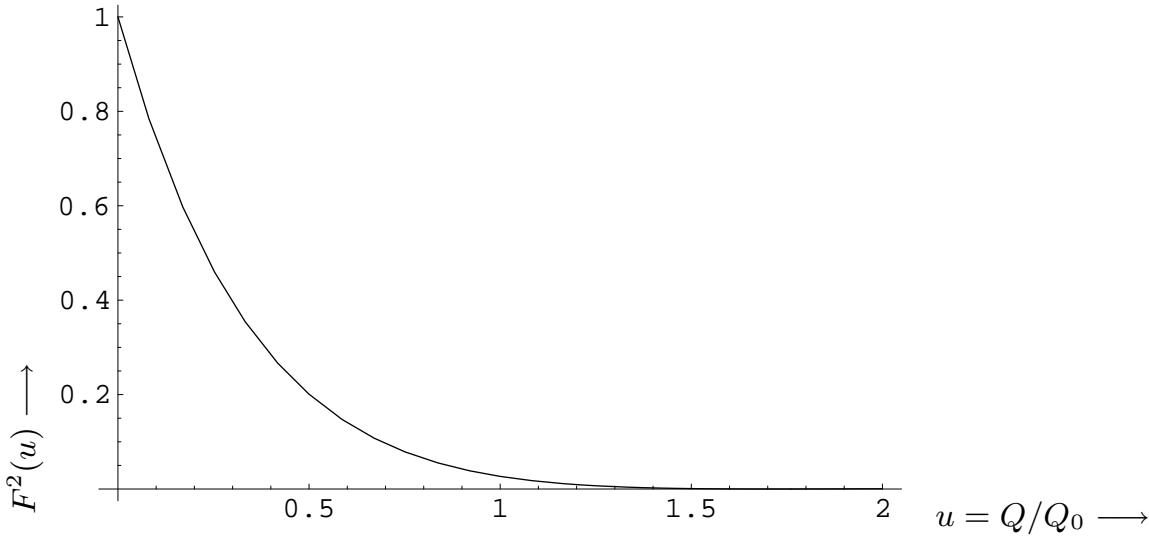}
 \caption{The form factor $F^2(u)$ employed in our calculation. $u$ is the energy transfer to the nucleus in units of $Q_0$, i.e. $u=Q/Q_0$, with $Q_0=64$ keV.}
 \label{sqformf}
 \end{center}
  \end{figure}

  The differential rate discussed in the previous section depends on the nucleus via its form factor and its mass. It also depends on the WIMP mass through the reduced mass $\mu_r$ entering the parameter $a$.  For our numerical study we will
  focus on $^{127}$I, which is one of the most popular targets employed.
  The nuclear form factor we use was obtained in the shell model description of the target and is shown in
 Fig. \ref{sqformf}.

\begin{figure}[!ht]
 \begin{center}
 \includegraphics[clip,width=1.0\linewidth]{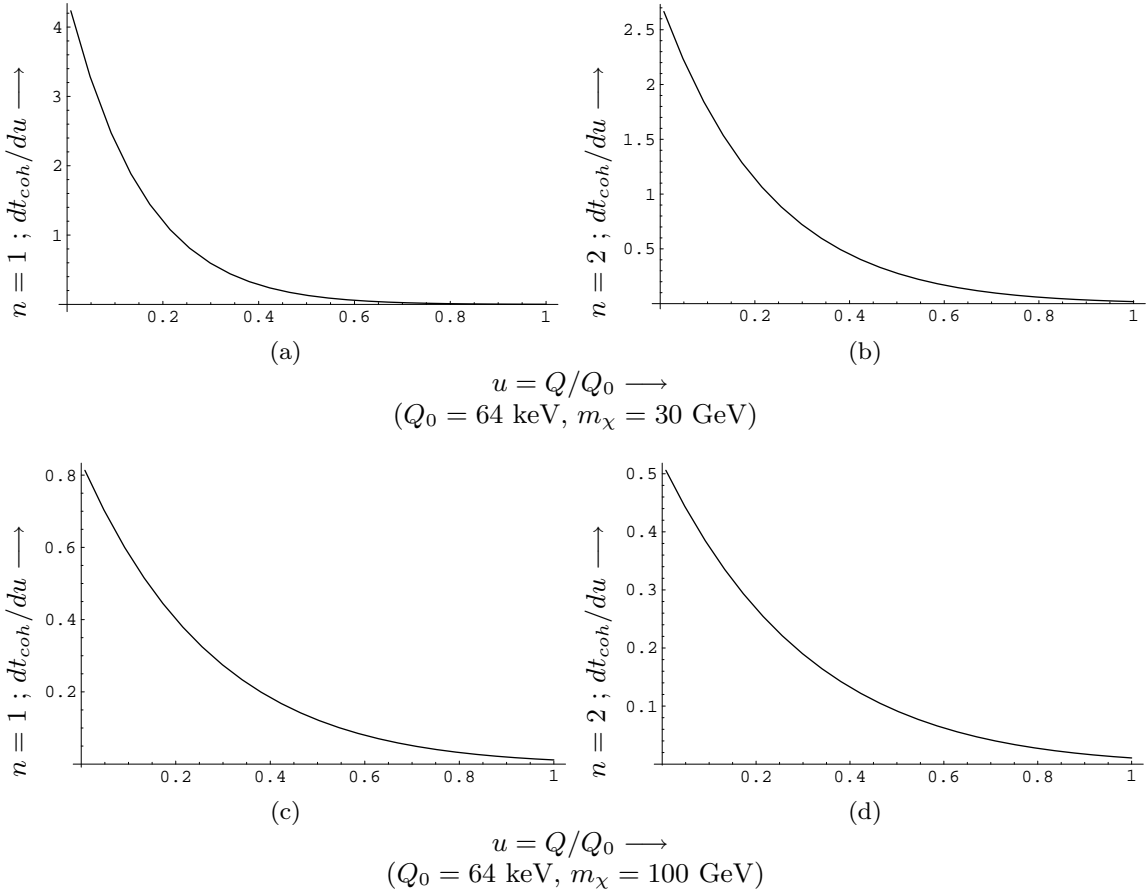}
 \caption{The quantity $ {dt_{coh}}/{du}$ for $n=1$ on the left and $n=2$ on the right. From top to bottom $m_{\chi}=30$ and $m_{\chi}=100$ GeV.}
 \label{fig:dift}
 \end{center}
  \end{figure}

   The  part of the differential rate associated with $\Psi_0$, indicated by  ${dt_{coh}}/{du}$, is shown in Fig. \ref{fig:dift} for two WIMP masses $m_{\chi}=30$ and $100$ GeV, and $n=1,2$.
%  \end{document}
  The explicit results of this figure confirm those derived by inspection of Fig. \ref{fig:nomod}.

The total (time averaged) rate is given by:
 \beq
 t_{coh}=\int_{u_{min}}^{u_{max}} \frac{dt_{coh}}{du} du,
 \label{tcoh}
 \eeq
 where
%$\frac{dt_{coh}}{du}$ is the time average of the expression $\frac{dr}{du}$ of Eq. \ref{eq:rrate} and
$u_{min}$ is determined by the detector threshold and
$u_{max}=(n y_{esc})^2/a^2$ by the maximum WIMP velocity.
%The factor $A^2$ coming from the coherent contribution of all nucleons has not been included.\\
By including  both $\Psi_0(a \sqrt{u})$ and $H(a \sqrt{u})$ we can cast the rate in the form:
\begin{eqnarray}
r_{coh}&=&t_{coh} \left(1+h_{coh} \cos{\alpha} \right)
\nonumber \\
h_{coh}&=&\frac{1}{t_{coh}} \int_{u_{min}}^{u_{max}} \frac{dh_{coh}}{du} du.
\end{eqnarray}

\begin{figure}[!ht]
 \begin{center}
 \includegraphics[clip,width=1.0\linewidth]{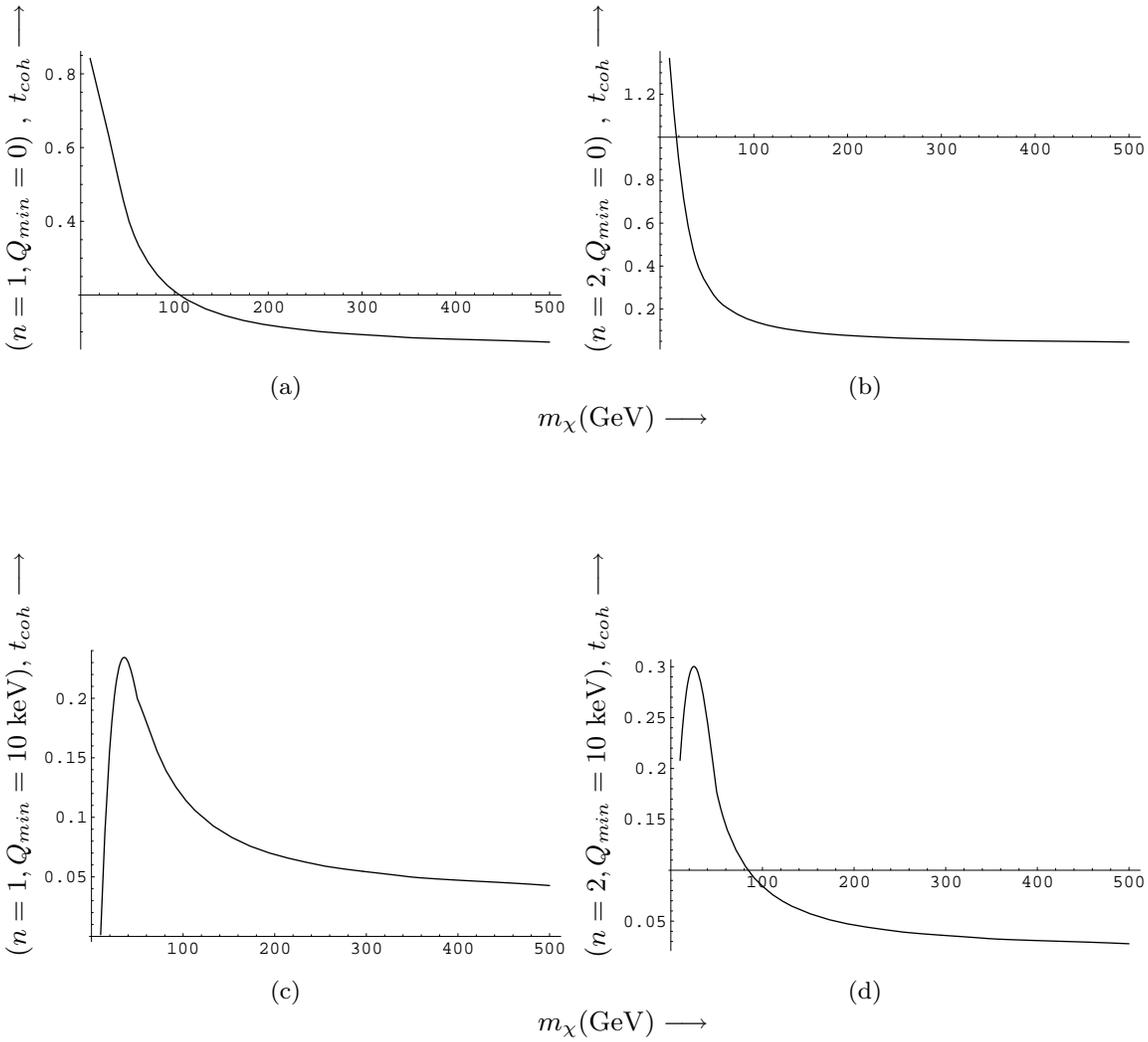}
 \caption{The quantity $t_{coh}$ for $Q_{min}=0$ at the top  and $Q_{min}=10$ keV at the bottom. From left to
right  $n=1$ and 2.}
 \label{fig:totalt}
 \end{center}
  \end{figure}

Integrating over the energy transfer, assuming either no detector cut off ($u_{min}=0$) or a cut off of $Q_{th}=10$ keV, we obtain the results shown in Fig. \ref{fig:totalt}.
  One can see from Fig. \ref{fig:totalt} that, except for the case of light WIMPs, the total rate is decreasing with increasing $n$. The reason is that, as we have seen in the previous section, even though the total area under the curves of Fig. \ref{fig:nomod} is independent of $n$, the nuclear form factor dumps out
  the high u components. Similarly, the area under the curves of Fig. \ref{fig:dift} decreases with $n$.
% This has a more dramatic effect in the case of high $n$.
% This mechanism is much more effective   in the case of small WIMP masses.
As expected, for a given $n$ the rate decreases as the energy cut off increases.
 \begin{figure}[!ht]
 \begin{center}
 \includegraphics[clip,width=1.0\linewidth]{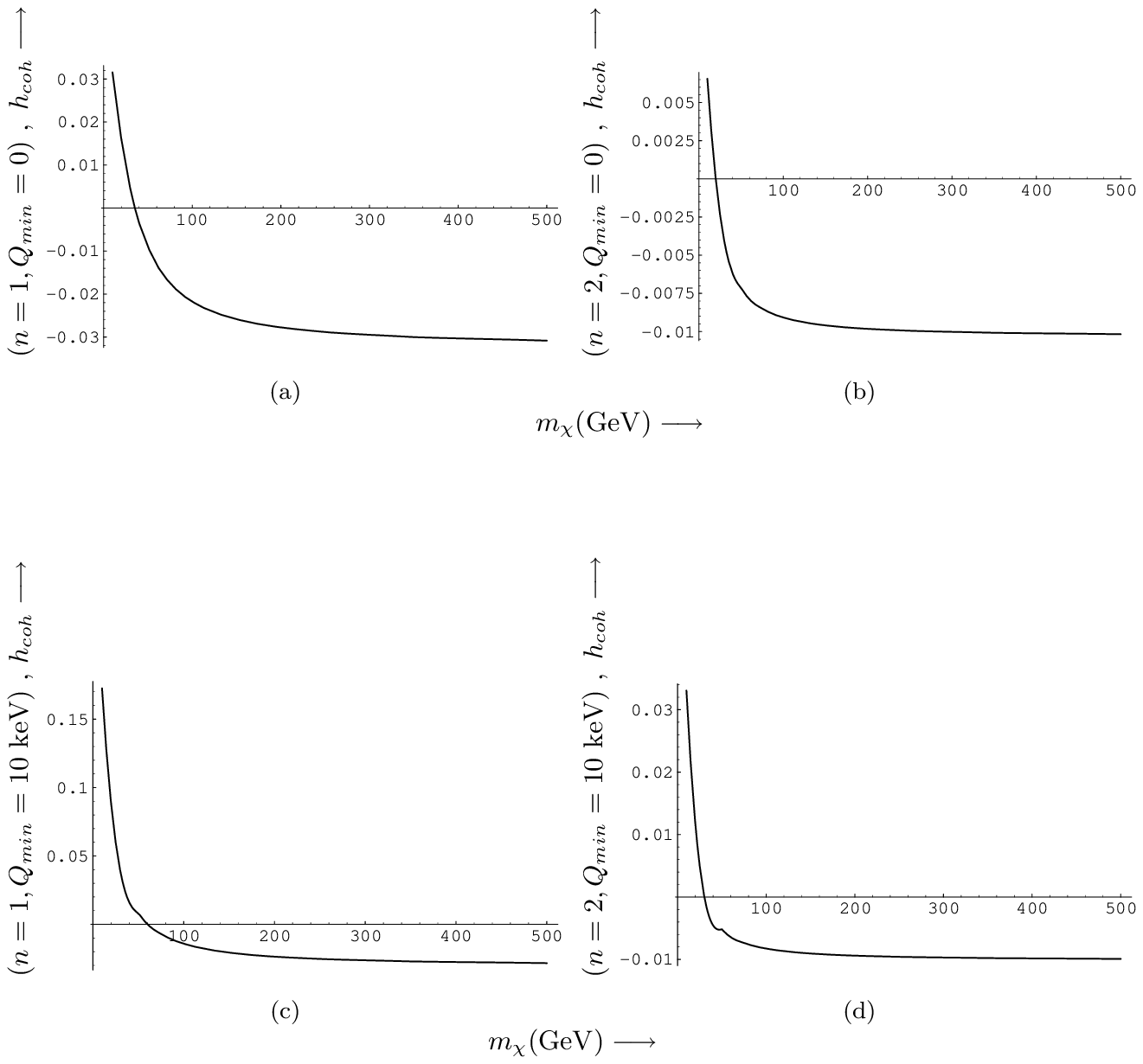}
 \caption{The quantity $h_{coh}$ for $Q_{min}=0$ at the top  and $Q_{min}=10$ keV at the bottom.
 From left to right $n=1$ and 2.}
 \label{fig:totalh}
 \end{center}
  \end{figure}
  From Fig. \ref{fig:totalt} we see that the modulation amplitude $h$ rapidly decreases as $n$ increases. This is expected
  in view of Fig. \ref{fig:mod}. For a given $n$ the modulation amplitude increases as the energy cut off increases. The reason
is that $h$ essentially is the ratio of the modulated amplitude divided by the unmodulated amplitude. Both decrease with increasing
$Q_{min}$, but the denominator is decreasing much more rapidly. In other words the increase in the modulation occurs at the
expense of the number of counts.

\section{Conclusions}

According to our present understanding of the evolution of the Universe, the main contribution to its energy
content comes from two sectors, the dark energy and dark matter, that have not been directly observed.
Within the majority of the models that have been proposed for the description of these sectors, there is no
coupling between them other than the gravitational one. In general, this very strong assumption is not
supported by some reasoning based on fundamental properties of the model, such as symmetries.
On the other hand, a coupling between the two sectors may even be desirable, as it may provide an
explanation of the coincidence problem, i.e. the comparable contributions of the two sectors
to the energy density of the Universe today.

The main motivation for this paper has been the wish to explore the direct
observational consequences of such a coupling.
We modeled the interaction between dark matter and dark energy
by assuming that the mass of the dark matter particles depends on the scalar field whose potential provides the
dark energy. The fact that the dark energy field, if it plays a dynamical role in the cosmological evolution today, must be effectively massless at length scales below the
horizon means that its presence results
in a long-range attractive force in the dark matter sector.
This can have significant implications for the mechanisms of
structure formation. The effect that has been of interest to us is the modification of the standard
isothermal Maxwell-Boltzmann distribution of dark matter in the galaxy halos \cite{tetradis,brouzakis}.
The main modification is that
the characteristic dark matter velocity can be increased significantly in the presence of the additional
force. As the velocity affects directly the detection rates of the various experiments that search for
dark matter, a detailed calculation of these rates, taking into account the new interaction, is important.

The modification of the velocity distribution has consequences for direct WIMP detection.
Regarding the (time averaged) event rates, our results depend on the WIMP mass.
For light WIMPs we find an increase of the rates  by about $50\%$, independently of the detector energy cutoff,
if the new force is stronger than the gravitational force by a factor $\kappa =\sqrt{3}$.
For larger masses, however, the new force leads to a substantial decrease in the rates. The
reason for this is that large energy transfers are inhibited by the nuclear form factor.
The consequences of the new interaction are more
pronounced in the case of the modulation amplitude. For light WIMP masses the modulation is decreased by an
order of magnitude for an ideal detector (zero energy threshold) for $\kappa=\sqrt{3}$ relative to
$\kappa=0$.
For heavy WIMPs the decrease is about a
factor of 4. The effect persists, but is somewhat less pronounced, in the case of a detector with a finite
energy threshold, e.g. about $10$ keV.

We should emphasize that the average WIMP energy, in addition to its linear increase with the WIMP mass, also
increases proportionally to $1+\kappa^2$ (see Eq. (\ref{avenergy})). The same holds for the
maximum WIMP energy, which depends on the escape velocity and scales with the same factor
(see Eq. (\ref{escvv})).
This provides the opportunity of planning novel
experiments other than those involving nuclear recoil, e.g. experiments detecting transitions to excited nuclear states in the MeV
region. The relevant rate may be enhanced significantly due to the tail of the velocity distribution.

\vspace {0.5cm}
\noindent{\bf Acknowledgments}\\
\noindent
This work was supported by
the RTN contract MRTN--CT--2004--503369 of the European Union. The work of N.T. was also supported by
the research program
``Pythagoras II'' (grant 70-03-7992)
of the Greek Ministry of National Education, partially funded by the
European Union,
and the research program ``Kapodistrias'' of the University of Athens.
Part of the work of one of the authors (J.D.V.) was performed during a visit to
T\"ubingen as a Humboldt awardee.

%\bibliography{TeX}
  \end{document}